\documentclass{article}
\usepackage[margin=1.5in]{geometry}
\usepackage[utf8x]{inputenc}
\usepackage{amsmath}
\usepackage{multirow,makecell}
\usepackage{url}
\usepackage{graphicx}
\usepackage{authblk}
\usepackage{listings}
\usepackage{tablefootnote}
\usepackage{threeparttable}

\title{Botometer 101: \\ Social bot practicum for computational social scientists}

\author[1]{Kai-Cheng Yang}
\author[2]{Emilio Ferrara}
\author[1]{Filippo Menczer}
\affil[1]{Observatory on Social Media, Indiana University Bloomington, USA}
\affil[2]{Information Sciences Institute, University of Southern California, USA}

\begin{document}

\newcommand{\code}[1]{{\texttt{#1}}}
\newcommand{\cashtag}[1]{{\$\texttt{#1}}}

\maketitle

\abstract{
Social bots have become an important component of online social media.
Deceptive bots, in particular, can manipulate online discussions of important issues ranging from elections to public health, threatening the constructive exchange of information.
Their ubiquity makes them an interesting research subject and requires researchers to properly handle them when conducting studies using social media data.
Therefore, it is important for researchers to gain access to bot detection tools that are reliable and easy to use.
This paper aims to provide an introductory tutorial of Botometer, a public tool for bot detection on Twitter, for readers who are new to this topic and may not be familiar with programming and machine learning.
We introduce how Botometer works, the different ways users can access it, and present a case study as a demonstration.
Readers can use the case study code as a template for their own research.
We also discuss recommended practice for using Botometer.
}

\section{Introduction}

Social bots are social media accounts controlled in part by software that can post content and interact with other accounts programmatically and possibly automatically~\cite{ferrara2016rise}.
While many social bots are benign, malicious bots can deceptively impersonate humans to manipulate and pollute the information ecosystem.  
Such malicious bots are involved with all types of online discussions, especially controversial ones.
Studies have identified interference of social bots in U.S. elections~\cite{shao_spread_2018,gorodnichenko_social_2021,bessi_social_2016,ferrara_characterizing_2020}, French elections~\cite{ferrara_disinformation_2017}, the Brexit referendum~\cite{bastos_public_2018,bastos_brexit_2019,gorodnichenko_social_2021,duh_collective_2018}, German elections~\cite{keller_social_2019}, and the 2017 Catalan referendum~\cite{stella_bots_2018}.
Bots also actively participate in public health debates~\cite{jamison_malicious_2019} including those about vaccines~\cite{broniatowski_weaponized_2018,yuan_examining_2019}, the COVID-19 pandemic~\cite{ferrara_what_2020,shi_social_2020,uyheng_bots_2020,yang2020prevalence}, and cannabis~\cite{allem_cannabis_2020}.
Research has also reported on the presence of social bots in discussions about climate change~\cite{marlow_twitter_2020,marlow_bots_2021,chen_social_2021}, cryptocurrency~\cite{nizzoli_charting_2020}, and the stock market~\cite{cresci_cashtag_2019,fan_social_2020}.

Malicious social bots demonstrate various behavioral patterns in their actions.
They may simply generate a large volume of posts to amplify certain narratives~\cite{marlow_bots_2021,keller2020political} or to manipulate the price of stocks~\cite{cresci_cashtag_2019,fan_social_2020} and cryptocurrencies~\cite{nizzoli_charting_2020}.
They can also disseminate low-credibility information strategically by getting involved in the early stage of the spreading process and targeting popular users through mentions and replies~\cite{shao_spread_2018}.
Some bots act as fake followers to inflate the popularity of other accounts~\cite{bilton_social_2014,confessore_follower_2018,varol2020journalists}.
In terms of content, malicious bots are found to engage other accounts with negative and inflammatory language~\cite{stella_bots_2018} or hate speech~\cite{albadi_hateful_2019,uyheng_bots_2020}.
In some cases, bots form dense social networks to boost engagement and popularity metrics and to amplify each other's messages~\cite{caldarelli_role_2020,torres2020trains,chen_neutral_2021}.

Most existing reports and studies on social bots focus on Twitter, largely because its data can be easily accessed.
Although Twitter strengthened their efforts to contain malicious actors in recent years,\footnote{\url{blog.twitter.com/common-thread/en/topics/stories/2021/the-secret-world-of-good-bots}} deceptive bots remain prevalent and display evolving tactics to evade detection~\cite{yang2019arming}.
This has two implications for researchers.
First, characterizing the behavior of and assessing the impact of social bots remains an interesting research topic~\cite{rahwan_machine_2019}.
Second, researchers need to properly handle bots in their data since their presence may distort analyses~\cite{jamison_malicious_2019,ledford_social_2020}. 
It is therefore crucial for researchers to have access to a reliable tool for detecting social bots.

This practicum aims to provide a tutorial for Botometer, a machine learning tool for bot detection on Twitter.
Although other bot detection tools such as tweetbotornot\footnote{An R package for classifying Twitter accounts as bot or not available at \url{github.com/mkearney/Tweetbotornot}} and Bot Sentinel\footnote{A platform that classifies and tracks inauthentic accounts and toxic trolls available at \url{botsentinel.com}} exist, we focus on Botometer for several reasons.
First, it is well maintained and has been serving the community for the past seven years without major outages.
It has also been routinely upgraded to stay accurate and relevant.
Second, Botometer is easily accessible through both a web interface and an application programming interface (API).
Anyone with a Twitter account can use the web version for free; researchers with Twitter developer accounts can use the API endpoints to analyze large-scale datasets. The API has a nominal fee for heavy use, which discourages abuse and partially offsets infrastructure and maintenance costs.
Third, Botometer is quite popular. It handles around a quarter million daily queries---over half a billion in total since its inception. 
Finally, Botometer has been extensively validated in the field.
Many researchers have applied Botometer in their studies to directly investigate social bots and their impact~\cite{broniatowski_weaponized_2018,keller_social_2019,allem_cannabis_2020,fan_social_2020}, or to distinguish human accounts and bot-like accounts in order to better address their questions of interest~\cite{vosoughi_spread_2018,grinberg_fake_2019,bovet_influence_2019}.

This tutorial is designed for data scientists and computational social scientists who might not be familiar with Botometer, the machine learning methods behind it, its programmatic interface, or how to interpret its results.
We start with an introduction to how Botometer works and how users can access it.
We then present a case study to demonstrate Botometer usage.
The source code for this case study is shared through a public repository for readers to replicate this analysis and use it as a template for their own research.
We finally discuss recommended practice.

\section{How Botometer works}

\begin{figure}
  \centering
  \includegraphics[width=\linewidth]{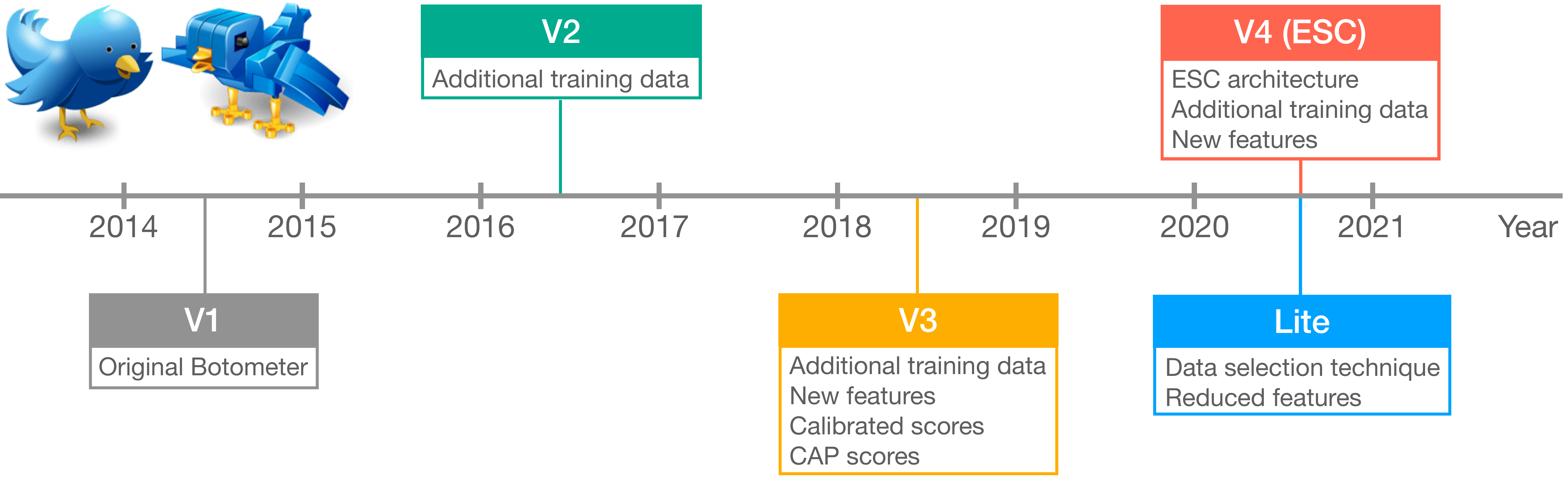}
  \caption{
  The timeline of Botometer versions.
  }
  \label{fig:botometer_history}
\end{figure}

Figure~\ref{fig:botometer_history} presents the timeline and key characteristics of successive Botometer versions over the years.  
Since the behaviors of bot and human accounts evolve over time, version upgrades are necessary for Botometer to stay accurate and relevant.
Upgrades typically included adding new training data and updating model features. The most recent version also involved major architectural changes.
Users of Botometer should be aware that results from different versions are usually not comparable and the format of input and output might change as well. 

For details of early versions such as V2~\cite{varol2017online} and V3~\cite{yang2019arming}, readers can refer to the corresponding papers.
This tutorial focuses on V4~\cite{sayyadiharikandeh2020detection}.
In addition to new training data and new features, this version introduced a new architecture. 
We will also briefly cover a recently added model for fast bot detection~\cite{yang_scalable_2020}.

\subsection{Supervised machine learning for bot detection}

Under the hood, Botometer is a supervised machine learning classifier that distinguishes bot-like and human-like accounts based on their features (i.e., characteristics). Unsupervised methods have also been proposed in the literature~\cite{chavoshi2016debot,echeverria2017discovery1}, but they only allow for the detection of specific, predefined behaviors. Therefore they are not suitable to build a general detection tool. 

Botometer considers over 1,000 features that can be categorized into six classes: user profile, friends, network, temporal, content and language, and sentiment~\cite{varol2017online}.
For example, the user profile category includes features such as the length of the screen name, whether the account uses the default profile picture and background, the age of the account, etc.
The content and language category consists of features such as the number of verbs, nouns, and adjectives in the tweets.
For a given account, these features are extracted and encoded as numbers.
This way the account can be represented by a vector of feature numbers, enabling machine learning classifiers to process the information.

\begin{table}
\caption{
Annotated datasets of human and bot accounts used to train Botometer.
}
\centering
\begin{tabular}{lrrll}
   \hline
   Dataset & Bots & Humans & Annotation method & Ref. \\
   \hline
   \texttt{varol-icwsm} & 733 & 1,495 & Human annotation & \cite{varol2017online} \\
   \texttt{cresci-17}  & 7,049 & 2,764 & Various methods & \cite{cresci2017paradigm} \\
   \texttt{pronbots} & 17,882 & 0 & Spam bots & \cite{yang2019arming} \\
   \texttt{celebrity}  & 0 & 5,918 & Celebrity accounts & \cite{yang2019arming} \\
   \texttt{vendor-purchased} & 1,087 & 0 & Fake followers &\cite{yang2019arming} \\
   \texttt{botometer-feedback} & 139 & 380 & Human annotation & \cite{yang2019arming} \\
   \texttt{political-bots} & 62 & 0 & Human annotation & \cite{yang2019arming} \\
   \texttt{gilani-17} & 1,090 & 1,413 & Human annotation  & \cite{gilani2017bots} \\
   \texttt{cresci-rtbust} & 353 & 340 & Human annotation &\cite{mazza_rtbust_2019} \\
   \texttt{cresci-stock} & 7,102 & 6,174 & Signs of coordination & \cite{cresci_fake_2018} \\
   \texttt{botwiki}  & 698 & 0 & Self-declared & \cite{yang_scalable_2020} \\
   \texttt{midterm-2018}  & 0 & 7,459 & Human annotation & \cite{yang_scalable_2020} \\
   \texttt{astroturf} & 505  & 0 & Human annotation & \cite{sayyadiharikandeh2020detection} \\
   \texttt{kaiser} & 875 & 499 & Politicians + bots  & \cite{harvardDataset} \\
   \hline
\end{tabular}
   \label{table:dataset} 
\end{table}

Supervised machine learning algorithms such as Botometer depend on the availability of training data---accounts labeled as either human or bot. These labels usually come from human annotation \cite{varol2017online}, automated methods (e.g., honey pots \cite{lee2011seven}), or botnets that display suspicious behaviors \cite{echeverria2017discovery2,echeverria2017discovery1}. A critical issue with existing datasets is the lack of ground truth. There is no objective, agreed-upon, operational definition of social bot. A further complicating factor is the prevalence of accounts that lie in the gray area between human and bot behavior, where even experienced researchers cannot easily discriminate. Nevertheless, datasets do include many typical bots; using the training labels as proxies for ground truth makes it possible to build practically viable tools.

Botometer-V4 is trained on a variety of datasets shown in Table~\ref{table:dataset}, which are publicly available in a Bot Repository.\footnote{\url{botometer.osome.iu.edu/bot-repository}}
With all training accounts being represented as feature vectors, a classifier can learn the characteristics of bot and human accounts.
Botometer uses a classification model called Random Forest, which consists of many rules learned from the training data. 

To evaluate a Twitter account, Botometer first fetches its 200 most recent tweets and tweets mentioning it from Twitter, extracts its features from the collected data, and represents this information as a feature vector.
Each model rule uses some of the features and provides a vote on whether an account is more similar to bot or human accounts in the training data. Based on how many rules vote for the bot or human class, the model provides a ``bot score'' between zero and one:   
a score close to one means the account is highly automated, while a score near zero means a human is likely handling the account.
Some accounts may demonstrate the characteristics of both humans and bots.
For instance, a bot creator might generate content like a regular user but uses a script to control many accounts. 
These cases can be confusing for the classifier, which would then produce scores around 0.5.

While human accounts tend to behave similarly, different types of bots usually have unique behavioral patterns. Based on this observation, Botometer-V4 uses several specialized Random Forest classifiers: one for each type of bots in the training data and one for humans. 
The results of this Ensemble of Specialized Classifiers (ESC) are aggregated to produce a final result.
More details about the ESC architecture can be found in the original paper~\cite{sayyadiharikandeh2020detection}.
At the end of the day, the ESC architecture is still a machine learning classifier, which yields scores between 0 and 1.
Different from a single Random Forest, the scores generated by ESC tend to have a bimodal distribution.

It is worth mentioning that the content and language features and sentiment features are based on English.
When a non-English account is passed to Botometer, these features become meaningless and might affect the classification.
As a workaround, Botometer also returns a language-independent score, which is generated without any language-related features.
Users need to be aware of the account language and choose the most appropriate Botometer score.

\subsection{Model accuracy}

The accuracy of the model is evaluated through 5-fold cross-validation on the annotated datasets shown in Table~\ref{table:dataset}.
Simply speaking, the classifier is trained on part of the annotated datasets and tested on the rest to provide a sense of its accuracy.
In the experimental environment, Botometer works really well. V4 has an AUC (area under the receiver operating characteristic curve) of 0.99, suggesting that the model can distinguish bot and human accounts in Table~\ref{table:dataset}---as well as accounts in the wild that resemble those in the training datasets---with very high accuracy.

However, Botometer is not perfect and may misclassify accounts due to several factors.
For example, the training datasets might have conflicts because they were created by different people with different standards.
In some cases Botometer fails to capture the features that can help distinguishing different accounts.
Botometer sometimes struggles with inactive accounts since not enough data is available for evaluation.
The accuracy of the model may further decay when dealing with new accounts different from those in the training datasets.
These accounts might come from a different context, use different languages other than English~\cite{rauchfleisch_false_2020,martini_bot_2021}, or show novel behavioral patterns~\cite{cresci2017paradigm,yang2019arming,Dimitriadis_multiclass_2021}.
These limitations are inevitable for all supervised machine learning algorithms, and are the reasons why Botometer has to be upgraded routinely.

Some critics exploit these limitations to undermine the entire 
field of study devoted to social bots. 
For example, one might select small sets of accounts with large false-positive error rates to argue that no bot detection tool is valid or that social bots do not exist at all. These arguments use fallacies such as cherry-picking and strawman in disingenuous ways. Validation through manual annotations is extremely valuable, especially when highlighting cases where existing machine learning models perform poorly, but should be used in constructive ways. New manually-annotated datasets should be made available, ideally via the public Bot Repository, to support the development of improved models.

\subsection{Results interpretation}

Early versions of Botometer returned to users raw scores in the unit interval, produced by the Random Forest classifiers.
Although users often treated them as probabilities, such interpretation is inaccurate.
Consider Twitter accounts $a$ and $b$ and their respective scores 0.7 and 0.3 produced by a Random Forest classifier.
We can say that $a$ is more bot-like than $b$, but it is inaccurate to say that there is a 70\% chance that $a$ is a bot or that $a$ is 70\% bot.
Since Botometer-V3, the scores displayed in the web interface are rescaled to the range 0--5 to discourage inaccurate probabilistic interpretations.

For users who need a probabilistic interpretation of a bot score, the Complete Automation Probability (CAP)  represents the probability that an account with a given score or greater is automated. CAP scores have also been available since Botometer-V3.
The CAP scores are Bayesian posteriors that reflect both the results from the classifier and prior knowledge of the prevalence of bots on Twitter, so as to balance false positives with false negatives.
For example, suppose an account has a raw bot score of 0.96/1 (equivalent to 4.8/5 display score on the website) and a CAP score of 90\%.
This means that 90\% of accounts with a raw bot score above 0.96 are labeled as bots, or, as indicated on the website, 10\% of accounts with a bot score above 4.8/5 are labeled as humans.
In other words, if you use a threshold of 0.96 on the raw bot score (or 4.8 on the display score) to classify accounts as human/bot, you would wrongly classify 10\% of accounts as bots---a false positive rate of 10\%. This helps researchers determine an appropriate threshold based on acceptable false positive and false negative error rates for a given analysis.

\subsection{Fast bot classification} 

When Botometer-V4 was released, a new model called BotometerLite was added to the Botometer family~\cite{yang_scalable_2020}.
BotometerLite was created to enable fast bot detection for large scale datasets.
The speed of bot detection methods is bounded by the platform's rate limits.
For example, the Twitter API endpoint used by Botometer-V4 to fetch an account's most recent 200 tweets and recent mentions from other users has a limit of 43,200 accounts per app key, per day.
Many studies using Twitter data have millions of accounts to analyze; with Botometer-V4, this may take weeks or even months.

To achieve scalability, BotometerLite relies only on features extracted from user metadata, contained in the so-called user object from the Twitter API.
The rate limit for fetching user objects is over 200 times the rate limit that bounds Botometer-V4. 
Moreover, each tweet collected from Twitter has an embedded user object.
This brings two extra advantages.
First, once tweets are collected, no extra queries to Twitter are needed for bot detection.
Second, the user object embedded in each tweet reflects the user profile at the moment when the tweet is collected.
This makes bot detection on archived historical data possible.

In addition to the improved scalability, BotometerLite  employs a novel data selection mechanism to ensure its accuracy and generalizability.
Instead of throwing all training data into the classifier, a subset is selected by optimizing three evaluation metrics: cross-validation accuracy on the training data, generalization to holdout datasets, and consistency with Botometer.
This mechanism was inspired by the observation that some datasets are contradictory to each other.
After evaluating the classifiers trained on all possible combinations of candidate training sets, the winning classifier only uses five out of eight datasets but performs well in terms of all evaluation metrics.

BotometerLite allows researchers to analyze large-volume streams of accounts in real time, while the limited training data may involve a compromise in accuracy on certain bot classes compared to Botometer-V4. 
In terms of how to choose between the two endpoints, we still recommend using Botometer-V4 when feasible since it analyzes more data and produces more detailed results.

\section{Botometer interface}

Although the machine learning model might seem complicated, the interface of Botometer is designed to be easy to use.
Botometer has a website and API endpoints with similar functionality.
The website\footnote{\url{botometer.org}} is handy for users who need to quickly check several accounts.
With a Twitter account, users can access the Botometer website from any web browsers, even on their mobile devices.
The website is straightforward to use: after authorizing Botometer to fetch Twitter data, users just need to type a Twitter handle of interest and click the ``Check user'' button.

The Botometer Pro API\footnote{\url{rapidapi.com/OSoMe/api/botometer-pro}} can be more useful for research since it allows to programmatically check accounts in bulk.
The API is hosted by RapidAPI, a platform that helps developers manage API rate limits and user subscriptions.
Using the Botometer API requires keys associated with a Twitter app, which can be obtained through Twitter's developer portal.\footnote{\url{developer.twitter.com}} One also needs a RapidAPI account and a subscription to one of the API usage plans.

When querying the API, users are responsible to send the required data (i.e., 200 most recent tweets by the account being checked and tweets mentioning this account) in a specified format through HTTPS requests.
The Botometer API will process the data and return the results.
While queries can be sent through any programming language, we recommend using Python and the official \code{botometer-python} package that we maintain.\footnote{\url{github.com/IUNetSci/botometer-python}}
The package can fetch data from Twitter, format the data, and query the API on behalf of the user with a few lines of code:

\lstset{
    showspaces=false,
    basicstyle=\ttfamily\footnotesize
}
\begin{lstlisting}
import botometer

bom = botometer.Botometer(
    rapidapi_key="XYZ",
    consumer_key="XYZ",
    consumer_secret="XYZ",
    access_token="XYZ",
    access_token_secret="XYZ"
    )
result = bom.check_account("@yang3kc")

print(f"Bot score={result['display_scores']['english']['overall']}/5")
print(f"CAP score={result['cap']['english']:.2f}")
\end{lstlisting}
BotometerLite is also available as an endpoint through the Botometer Pro APIs.
We list the the input, output, and limitations of the API endpoints for Botometer-V4 and BotometerLite side by side in Table~\ref{table:api_specs}. 
We also summarize the common resources for using Botometer in Table~\ref{table:links} to help the readers navigate these resources.

\begin{table}
\centering
\caption{
Comparison of Botometer-V4 and BotometerLite APIs.
}
\begin{threeparttable}
\centering
\begin{tabular}{p{3cm}||p{3.0cm}|p{4.5cm}}
    \hline 
    Model & Botometer-V4 & BotometerLite \\
    \hline 
    Endpoint & Check account & Check account in bulk \\
    \hline 
    Query payload & User object, 200 most recent tweets, mentions & List of user objects and timestamps \\
    \hline 
    Response & Raw bot scores, sub-scores, CAP scores, basic account information, etc. & BotometerLite scores \\
    \hline
    Daily number of accounts allowed\tnote{*} & 43,200 & $\sim 8.6$ million \\
    \hline 
    Corresponding \code{botometer-python} method(s) & \code{check\_account} & \code{check\_accounts\_from\_tweets}, \code{check\_accounts\_from\_user\_ids}, \code{check\_accounts\_from\_screen\_names} \\
   \hline
\end{tabular}
\begin{tablenotes}
\item[*] The values represent the upper bounds based on Twitter's API rate limit when using a single app key. The actually numbers depend on other factors such as internet speed as well.
\end{tablenotes}
\end{threeparttable}
   \label{table:api_specs} 
\end{table}

\begin{table}
\caption{
Common resources for using Botometer.
}
\centering
\begin{tabular}{p{3cm}|p{3.5cm}|p{4cm}}
   \hline
   Resource name & Resource & Note \\
   \hline
   Botometer website & \url{botometer.org} & Web interface of Botometer: useful for checking a small amount of accounts \\
   \hline
   Botometer Pro API & \url{rapidapi.com/OSoMe/api/botometer-pro} & API of Botometer: useful for checking accounts in bulk programmatically \\
   \hline
   \code{Botometer-python} package & \url{github.com/IUNetSci/botometer-python} & Python package to access Botometer Pro API \\
   \hline
   Botometer case study & \url{github.com/osome-iu/Botometer101} & Case study using Botometer with source code\\
   \hline
   Bot repository & \url{botometer.osome.iu.edu/bot-repository} & Annotated training datasets for Botometer \\
   \hline
\end{tabular}
   \label{table:links} 
\end{table}

Note that both Botometer and Twitter APIs have rate limits, meaning that users can only make a certain number of queries in a given time period.
Please check the respective websites for detailed documentation.
Getting familiar with the rate limits can help researchers better estimate the time needed for their analysis.

\section{Case study}

Since some readers may not be familiar with programming, querying the API could be challenging.
Moreover, analyzing the results returned by Botometer API is not trivial.
In this section, we provide a simple case study as a demonstration.
Different ways of analyzing the data are shown with recommended practice.
We share the code for this case study in a public repository\footnote{\url{github.com/osome-iu/Botometer101}} so that readers can use it as a template for their own research.
Next we outline the data collection and analysis steps implemented in this software repository.

\subsection{Data collection}

Let us consider two cryptocurrency cashtags, \cashtag{FLOKI} and \cashtag{SHIB}, and the cashtag of Apple Inc., \cashtag{AAPL}, and attempt to quantify which is more amplified by bot-like accounts.
A cashtag works like a hashtag but consists of a dollar sign ``\$'' and a stock or cryptocurrency symbol to help users track related discussions. 
We use \code{Tweepy},\footnote{\url{tweepy.org}} a Python package that helps access the Twitter API, to search tweets containing these cashtags.
For each cashtag, we only collect 2,000 tweets, which are sufficient for the demonstration.

\begin{table}
\caption{
Numbers of tweets and unique accounts mentioning different cashtags in raw data and analytical sample.
}
\centering
\begin{tabular}{l|r|r|r|r}
   \hline
                    & \multicolumn{2}{c|}{Raw data}       &  \multicolumn{2}{c}{Analytical sample}    \\
   \hline
   Cashtag          & Tweets      & Unique accounts      & Tweets             & Unique accounts      \\
   \hline
   \cashtag{SHIB}   & 2,000       & 1,241                & 1,819              & 1,111                \\
   \cashtag{FLOKI}  & 2,000       & 937                  & 1,893              & 860                  \\
   \cashtag{AAPL}   & 2,000       & 1,107                & 1,864              & 1,006                \\
   \hline
\end{tabular}
   \label{table:number_of_tweets_accounts} 
\end{table}

First, let us count the number of unique accounts in each dataset, as shown in Table~\ref{table:number_of_tweets_accounts}.
The number of unique accounts is much smaller than the number of tweets in all three datasets, suggesting that some accounts tweeted the same cashtag multiple times.

The next step is to query the Botometer API for bot analysis.
Instead of going through each tweet and check every user encountered, researchers can keep a record of accounts already queried to avoid repetition and increase efficiency.
The Botometer API returns rich information about each account.
We recommend storing the full results from Botometer for flexibility.

\begin{figure}
  \centering
  \includegraphics[width=0.5\linewidth]{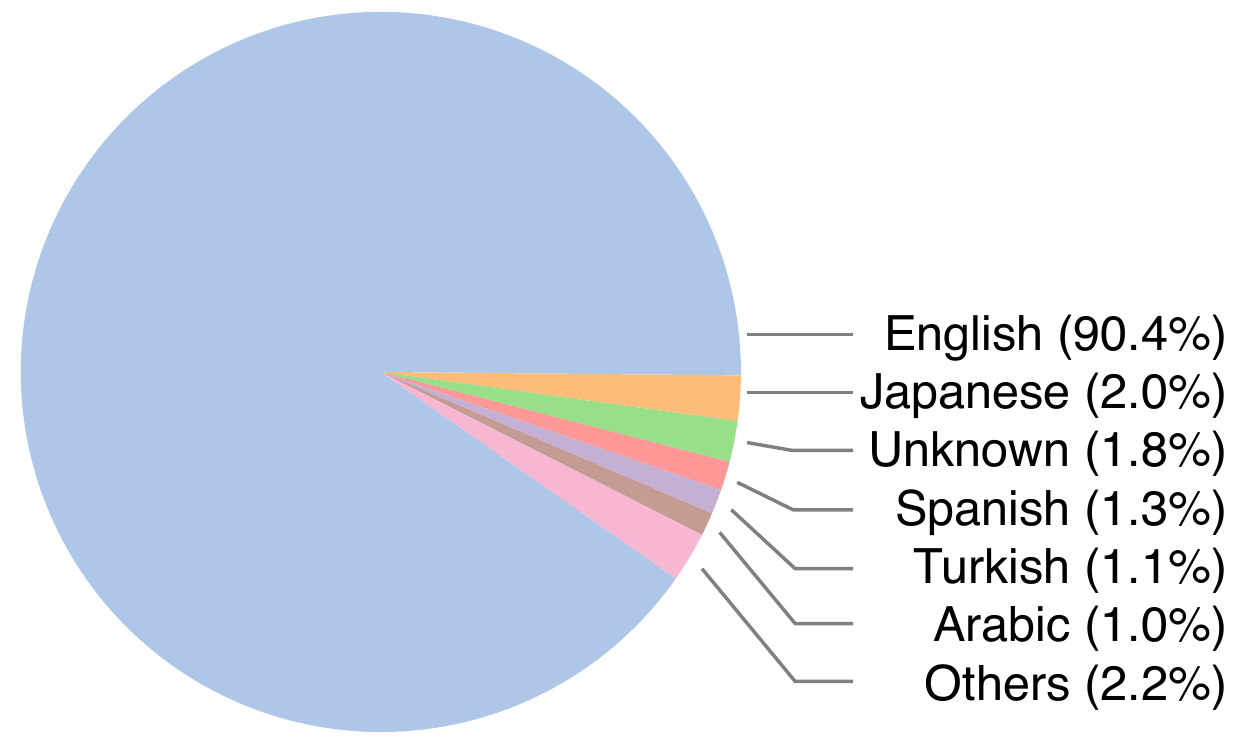}
  \caption{
  Percentage of accounts using each language in the three datasets combined.
  }
  \label{fig:lang_frequency}
\end{figure}

As mentioned above, Botometer generates an overall score and a language-independent score.
Since the two scores come from different classifiers, they are not comparable and should not be mixed together. To decide which one to use, let us calculate the proportion of accounts using each language. We can see in Figure~\ref{fig:lang_frequency} that the majority of accounts in our raw data tweet in English.
Therefore we only include English-speaking accounts and their tweets in our analytical sample (see Table~\ref{table:number_of_tweets_accounts} for summary statistics) and use the overall bot score.

\subsection{Analysis}

\begin{figure}
  \centering
  \includegraphics[width=1\linewidth]{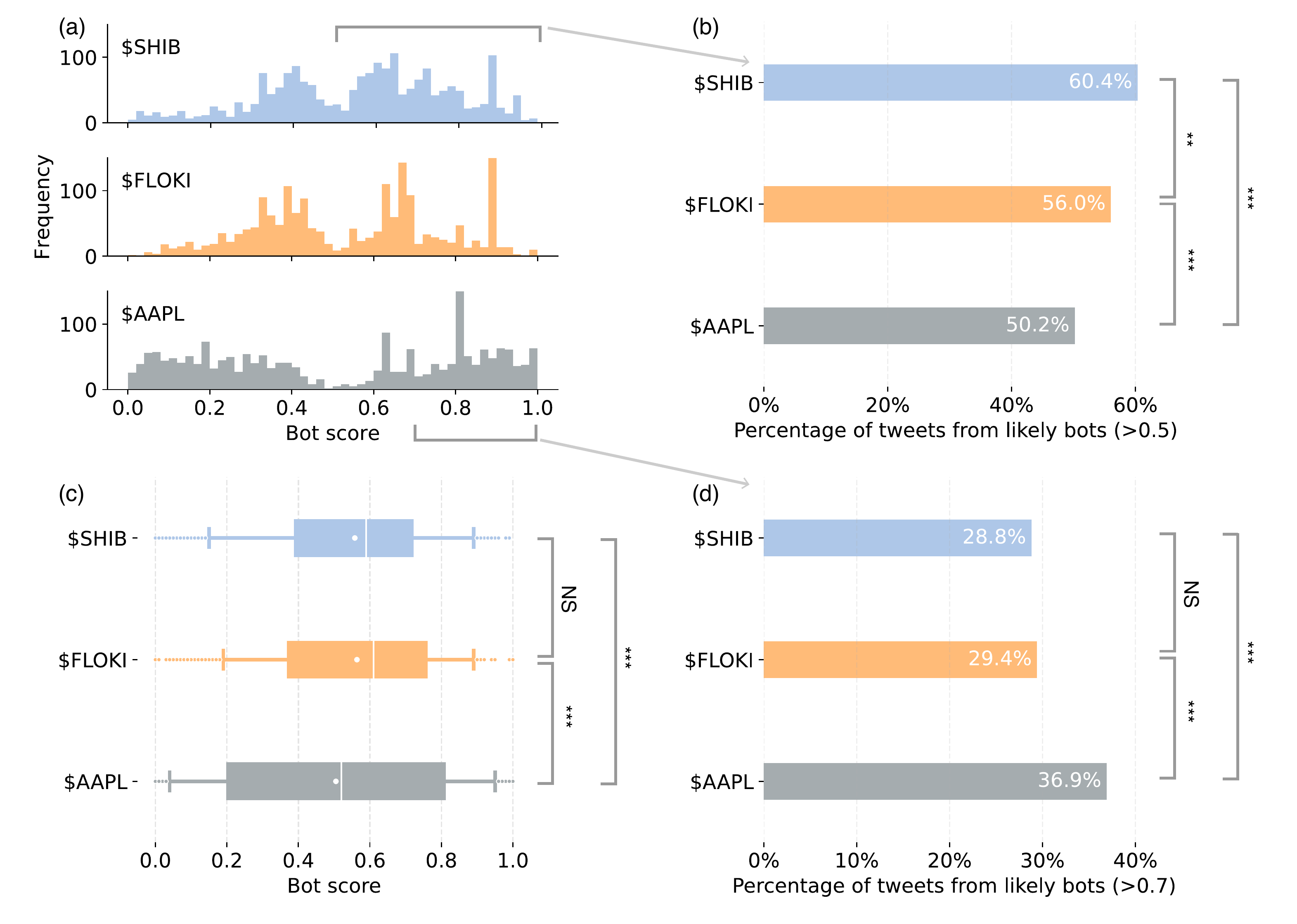}
  \caption{
  (a) Bot score distributions for tweets mentioning different cashtags.
  (b) Percentage of tweets posted by likely bots using 0.5 as a threshold. 
  (c) Box plots of the bot scores for tweets mentioning different cashtags.
  The white lines indicate the median values; the white dots indicate the mean values.
  (d) Similar to (b) but using a bot score threshold of 0.7.
  Statistical tests are performed for pairs of results in (b--d).
  Significance level is represented by the stars: ***$p \le 0.001$, **$p \le 0.01$, *$p \le 0.05$, NS$=p>0.05$.
  }
  \label{fig:bot_score_distribution}
\end{figure}

We plot the bot score distribution for tweets mentioning each cashtag in Figure~\ref{fig:bot_score_distribution}(a). 
Here we base our analysis on the raw scores in the unit interval.
Since we are interested in the bot activity level of each cashtag, we use tweets (as opposed to accounts) as the units of analysis.
This means that accounts tweeting the same cashtag multiple times have a larger contribution.

\begin{figure}
  \centering
  \includegraphics[width=0.45\linewidth]{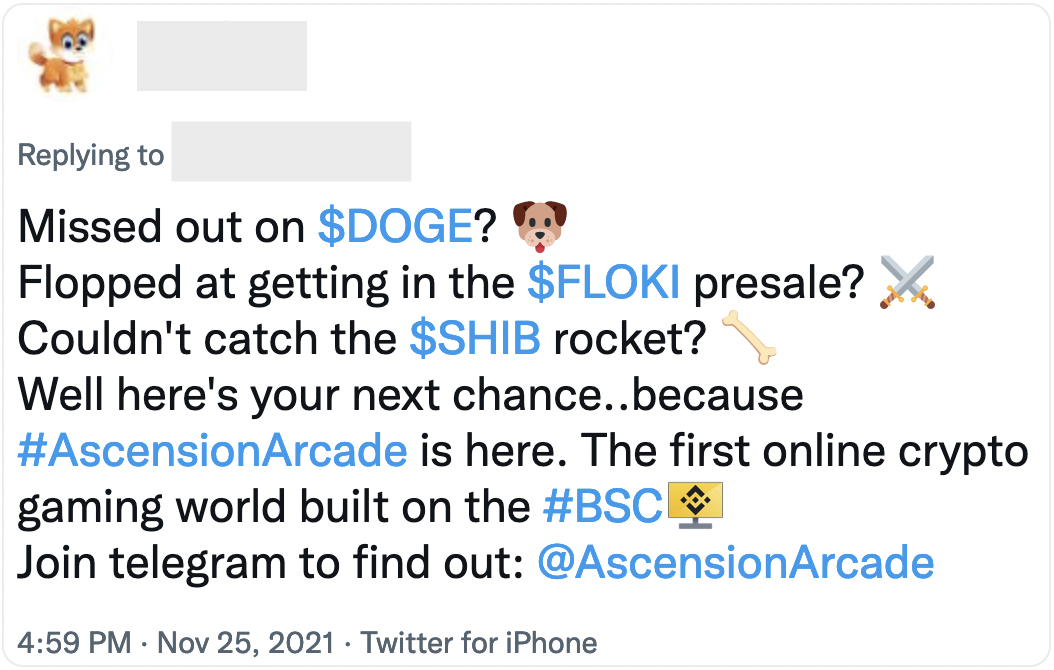}
  \caption{
  Screenshot of a bot-like account replying to a tweet containing the keyword ``NFT'' with a message promoting cryptocurrencies.
  The same message was replied by this account to a large number of tweets.
  }
  \label{fig:bot_screen_shot}
\end{figure}

In all three cases, the distribution has a bimodal pattern, a result of the ESC architecture of Botometer-V4.
We can observe some spikes in all cases, which are caused by accounts tweeting the same cashtag repeatedly.
For example, the spike near 0.89 for \cashtag{SHIB} and \cashtag{FLOKI} comes from a bot-like account that replied the same message promoting cryptocurrency tokens to a large number of tweets containing the keyword ``NFT''; see the screenshot of the message in Figure~\ref{fig:bot_screen_shot}.

To address our research question, we need to quantify the bot activity level for each cashtag and compare them.
The first approach is to compare their bot score distributions with two-sided Mann–Whitney U tests (see results in Figure~\ref{fig:bot_score_distribution}(c)).
The bot score distributions of \cashtag{SHIB} and \cashtag{FLOKI} are not significantly different from each other ($p=0.56$), but both of them have a higher bot activity level than \cashtag{AAPL} (\cashtag{SHIB} vs. \cashtag{AAPL}: $p<0.001$; \cashtag{FLOKI} vs. \cashtag{AAPL}: $p<0.001$).

The second approach dichotomizes the bot scores and considers the accounts with scores higher than a threshold as likely bots.
Then the proportion of tweets from likely bots can be calculated and compared.
In this approach, a threshold has to be chosen.
In the literature, 0.5 is the most common choice~\cite{shao_spread_2018,vosoughi_spread_2018,bessi_social_2016}; higher values, such as 0.7~\cite{grinberg_fake_2019} and 0.8~\cite{broniatowski_weaponized_2018}, are also used.
One may also consider running the same analysis with different threshold values to test the robustness of the findings~\cite{shao_spread_2018}.

Here we use both 0.5 and 0.7 as  thresholds and show the results in Figure~\ref{fig:bot_score_distribution}(b) and (d), respectively.
We apply two-proportions $z$-tests to estimate the significance level of the differences.
When using 0.5 as the threshold, the percentage of tweets from likely bots that mentioned \cashtag{SHIB} is significantly higher than those in the \cashtag{FLOKI} ($p=0.009$) and \cashtag{AAPL} datasets ($p<0.001$).
The percentage of tweets from likely bots that mentioned \cashtag{FLOKI} is also significantly higher than that in the \cashtag{AAPL} dataset ($p<0.001$).
However, when using 0.7 as the threshold, the results change: percentages of tweets from likely bots in \cashtag{SHIB} and \cashtag{FLOKI} datasets are no longer significantly different from each other ($p=0.38$); both of them are lower than that in the \cashtag{AAPL} dataset (\cashtag{SHIB} vs. \cashtag{AAPL}: $p<0.001$; \cashtag{FLOKI} vs. \cashtag{AAPL}: $p<0.001$).

In other studies, different approaches or threshold choices may yield consistent results.
However, they lead to seemingly different conclusions in this case. 
This is because different measures represent different properties of the bot score distribution.
If we revisit Figure~\ref{fig:bot_score_distribution}(a), we can see that although the distributions of \cashtag{SHIB} and \cashtag{FLOKI} scores have more mass in the $(0.5, 1]$ region than that of \cashtag{AAPL} scores, the mass tends to concentrate around 0.6, while the distribution of \cashtag{AAPL} scores has more mass near 1.
This nuanced difference causes the contradictory results when using different threshold values.

By reconciling the results from different approaches, we can answer our research question now. 
It appears that discussions about the cryptocurrencies \cashtag{SHIB} and \cashtag{FLOKI} show more automated activities than that about \cashtag{AAPL}, but among the accounts tweeting \cashtag{AAPL}, we find more highly automated bot-like accounts.
Note that the analysis here is mainly for demonstrating the use of Botometer; the samples of tweets analyzed are small and not representative of the entire discussion, so the conclusions only reflect the status of the collected data and should not be generalized.

\section{Recommended practice}

The sections above cover some recommended practice such as being careful when interpreting raw bot scores, being mindful about user language, and being aware of different versions of Botometer.
Here we make a few more recommendations to help avoid common pitfalls.

\subsection{Transient nature of Botometer scores}

\begin{figure}
  \centering
  \includegraphics[width=1\linewidth]{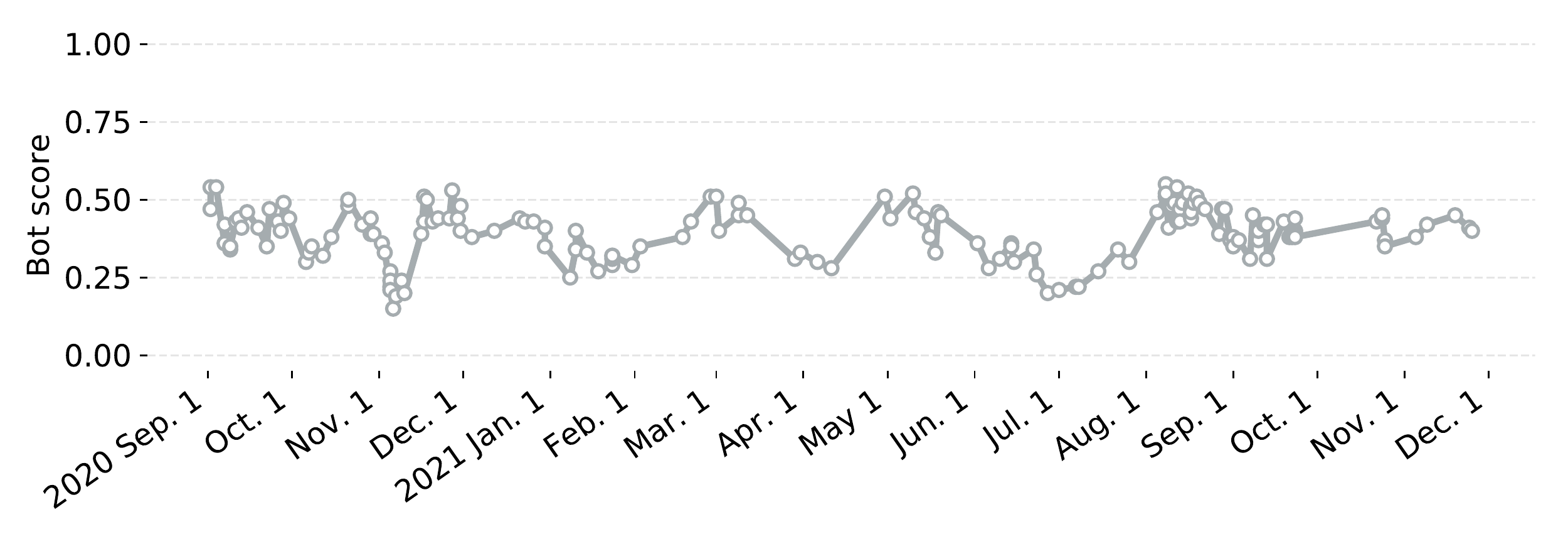}
  \caption{
  Time series of bot scores of an account from September 2020 to November 2021.
  The queries were not made regularly, so the time intervals between consecutive data points vary.
  }
  \label{fig:bot_score_change}
\end{figure}

Recall that Botometer uses the 200 most recent tweets by an account and other tweets mentioning the account for analysis.
This means that the results of Botometer change over time, especially for very active accounts.
To demonstrate this, we plot the time series of the overall bot score of an account in Figure~\ref{fig:bot_score_change}. 
This account posts roughly 16 tweets each week and gets mentioned by others frequently.
We can see that the bot score fluctuates over time.
In some other cases, an account might be suspended or removed after a while, making it impossible to analyze. 

Due to the transient nature of Botometer scores, a single bot score only reflects the status of the account at the moment when it is evaluated.
Users should be careful when drawing conclusions based on the bot scores of individual accounts.
For researchers, a common practice is to collect tweets first, then perform bot detection later.
To reduce the effect of unavailable accounts and to keep the bot scores relevant, bot analysis should be conducted right after data collection.

\subsection{Evaluating bot score distributions}

Whenever possible, we recommend collecting large datasets and use statistical analyses to evaluate bot activity based on comparisons of score distributions across different groups of accounts.
As demonstrated in the case study, bot score distributions can reveal rich information about the data.
Using distributions for analysis also reduces the uncertainty level of Botometer due to its imperfection and transient nature.
Most importantly, comparing distributions of scores---e.g., for accounts tweeting about a given topic versus a suitable baseline---allows for statistical tests that are impossible at the level of individual accounts.

\subsection{Validating thresholds}

In some analyses, dichotomizing the bot scores based on a threshold is necessary.
In these cases, we recommend validating the choice of threshold.
For researchers with the ability and resources, the ideal approach is to manually annotate a batch of bot and human accounts in their datasets. 
Such a preliminary analysis could be used, first, to determine whether Botometer is a helpful tool to evaluate a given scenario. Assuming it is, one can then vary the threshold and select the value that optimizes some appropriate metric on the annotated accounts.
Depending on the desire to maximize accuracy, minimize false positive errors, minimize false negative errors, or some combination, one can use metrics such as accuracy, precision, recall, or F1. 
When annotating additional accounts is not feasible, we suggest running multiple analyses using different threshold choices to confirm the robustness of the findings.

\subsection{Using Botometer in a civil way}

We have noticed that Botometer has been used to attack others.
For example, some users may call others with whom they disagree ``bots'' and use the results of Botometer as justification. 
This is a misuse of Botometer. Users should keep in mind that any classifier such as Botometer can mislabel individual accounts. Furthermore, even if an account is automated, it does not mean it is deceptive or malicious. 
Most importantly, such name calling is not helpful for creating healthy and informative conversations.  

\bibliographystyle{unsrt}
\bibliography{ref}

\end{document}